%% file: main.tex
\newif\iflncs
\lncsfalse

\iflncs
\documentclass[orivec]{llncs} 
\else
\documentclass[12pt]{article}
\usepackage[left=3cm,right=3cm]{geometry}
\fi

\PassOptionsToPackage{svgnames,x11names,dvipsnames}{xcolor}
\usepackage{xcolor}
\usepackage{tikz}
\usetikzlibrary{patterns}

\usepackage{color}
\usepackage{array,tabularx}
\usepackage{paralist}
\usepackage{enumerate}
\usepackage[algoruled,vlined,english]{algorithm2e}
\usepackage[color=green!20]{todonotes}
\usepackage[breaklinks]{hyperref}
\usepackage{url}
\usepackage{xspace}
\usepackage[draft]{fixme}
\usepackage{verbatim}
\usepackage{graphicx}
\usepackage{amssymb}
\usepackage{amsmath} 
\usepackage{listings}

\usepackage{authblk} 

\iflncs
\spnewtheorem{obs}{Observation}{\bfseries}{\itshape}
\spnewtheorem{prop}{Proposition}{\bfseries}{\itshape}
\spnewtheorem{corol}{Corollary}{\bfseries}{\itshape}
\else
\usepackage{amsthm}
\newtheorem{theorem}{THEOREM}[section]

\newtheorem{lemma}[theorem]{LEMMA}

\theoremstyle{definition}
\newtheorem{definition}[theorem]{Definition}

\newtheorem{example}[theorem]{EXAMPLE}

\newtheorem{obs}[theorem]{Observation}
\newtheorem{corol}[theorem]{Corollary}
\fi

\let\com=\newcommand
\com{\bthm}{\begin{theorem}}
\com{\ethm}{\end{theorem}}
\com{\bdfn}{\begin{definition}}
\com{\edfn}{\end{definition}}
\com{\blem}{\begin{lemma}}
\com{\elem}{\end{lemma}}
\com{\bcor}{\begin{corol}}
\com{\ecor}{\end{corol}}
\com{\bexm}{\begin{example}}
\com{\eexm}{\end{example}}
\com{\bobs}{\begin{obs}}
\com{\eobs}{\end{obs}}

\com{\bprf}{\begin{proof}}
\iflncs
\com{\eprf}{\qed\end{proof}}
\else
\com{\eprf}{\end{proof}}
\fi

\newcommand{\myref}[2]{\hyperref[#1]{\arabic{#1}$_{#2}$}}

\let\eps=\varepsilon
\let\vect=\vec
\renewcommand{\vec}[1]{\mathbf{#1}}
\newcommand{\tuple}[1]{\langle #1 \rangle}
\newcommand{\cvz}[0]{\mathbf{0}}  

\newcommand{\ints}{\ensuremath{\mathbb Z}\xspace}

\newcommand{\rats}{\ensuremath{\mathbb Q}\xspace}

\newcommand{\convhull}[0]{\mathrm{conv.hull}}
\newcommand{\cone}[0]{\mathrm{cone}}

\newcommand{\transitions}{\poly{Q}}

\newcommand{\trcv}[2]{\ensuremath{\bigl(\begin{smallmatrix}{#1}\hfill\\{#2}\hfill\end{smallmatrix}\bigr)}}

\let\tr=\trcv

\newcommand{\rfcoeff}[0]{a}

\newcommand{\poly}[1]{{\mathcal #1}}
\newcommand{\inthull}[1]{{#1}_I}
\newcommand{\intpoly}[1]{I({#1})}

\newcommand{\slc}[0]{\ensuremath{\mathit{SLC}}\xspace}

\newcommand{\lrf}[0]{\ensuremath{\mathit{LRF}}\xspace}
\newcommand{\mlrf}[0]{\ensuremath{\mathit{M\Phi}}RF\xspace}
\newcommand{\mlrfs}[0]{\ensuremath{\mathit{M\Phi}}RFs\xspace}

\newcommand{\lrfs}[0]{\ensuremath{\mathit{LRFs}}\xspace}
\newcommand{\llrf}[0]{\ensuremath{\mathit{LLRF}}\xspace}
\newcommand{\llrfs}[0]{\ensuremath{\mathit{LLRFs}}\xspace}
\newcommand{\bmsllrf}[0]{\llrf}
\newcommand{\bmsllrfs}[0]{\llrfs}
\newcommand{\mlrfq}[0]{\mbox{EM$\Phi$}RF\ensuremath{(\rats)}\xspace}
\newcommand{\mlrfz}[0]{\mbox{EM$\Phi$}RF\ensuremath{(\ints)}\xspace}
\newcommand{\dmlrfq}[0]{\mbox{BM$\Phi$}RF\ensuremath{(\rats)}\xspace}
\newcommand{\dmlrfz}[0]{\mbox{BM$\Phi$}RF\ensuremath{(\ints)}\xspace}

\newcommand{\trans}[0]{{\mbox{\tiny T}}}

\newcommand{\lp}[0]{\ensuremath{\mathit{LP}}\xspace}

\newcommand{\while}[0]{\ensuremath{\mathtt{while}\xspace}}
\newcommand{\wdo}[0]{\ensuremath{\mathtt{do}}\xspace}
\newcommand{\llrfsym}[0]{\ensuremath{\tau}\xspace}
\newcommand{\mlrfsym}[0]{\ensuremath{\tau}\xspace}

\newcommand{\diff}[1]{\ensuremath{\Delta #1}}

\newcommand{\ptime}[0]{\ensuremath{\mathtt{PTIME}}\xspace}
\newcommand{\coNP}[0]{\ensuremath{\mathtt{coNP}}\xspace}
\newcommand{\NP}[0]{\ensuremath{\mathtt{NP}}\xspace}

\newcommand{\ccone}[0]{{\mathtt{rec.cone}}}

\newcommand{\highlight}[2]{\begingroup\setlength{\fboxsep}{0.5pt}\colorbox{#1}{#2}\endgroup\xspace}

\title{On Multiphase-Linear Ranking Functions}
\iflncs
\author{Amir M. Ben-Amram\inst{1} \and Samir Genaim\inst{2}
} 
\institute{
School of Computer Science, The Tel-Aviv Academic College, Israel\\
\and
DSIC, Complutense University of Madrid (UCM),  Spain
}
\else
\author[1]{Amir M. Ben-Amram\thanks{amirben@mta.ac.il}}
\author[2]{Samir Genaim\thanks{samir.genaim@fdi.ucm.es}}
\affil[1]{The Academic College of Tel-Aviv Yaffo}
\affil[2]{Complutense University of Madrid}
\fi

\begin{document}
\maketitle

\input introduction.tex
\input preliminaries.tex

\input nested.tex

\input mlrf-vs-bms.tex

\input integer-mlrfs-bmsllrfs.tex
\input depth-bound.tex
\input iter-bound.tex
\input conclusions.tex

\bibliographystyle{plain}

\end{document}

%% file: introduction.tex

\begin{abstract}
Multiphase ranking functions (\mlrfs) were proposed as a means to prove the termination of a loop in which
the computation progresses through a number of ``phases", and the progress of each phase is described by a different linear ranking function.
Our work provides new insights regarding such functions for loops described by a conjunction of linear constraints (single-path loops).
We provide a complete polynomial-time solution to the problem
of existence and of synthesis of \mlrf of bounded depth (number of phases), when variables range over rational or real numbers;
a complete solution for the (harder) case that variables are integer, with a matching lower-bound proof, showing that the problem is coNP-complete;
and a new theorem which bounds the number of iterations for loops with \mlrfs. Surprisingly, the bound is linear,
even when the variables involved change in non-linear way. 
We also consider a type of \emph{lexicographic} ranking functions, \bmsllrf, more expressive than types of lexicographic functions for which
complete solutions have been given so far.  We prove that for  the above type of loops, lexicographic functions can be reduced to \mlrfs, and thus
the questions of complexity of detection and synthesis, and of resulting iteration bounds, are also answered for this class.
\end{abstract}

\section{Introduction}
\label{sec:intro}

Proving that a program will not go into an infinite loop is one of the
most fundamental tasks of program verification, and has been the
subject of voluminous research. Perhaps the best known, and often
used, technique for proving termination is the \emph{ranking
  function}.
This is a function $f$ that maps  program states into the
elements of a well-founded ordered set, such that $f(s) > f(s')$
holds whenever states $s'$ follows state $s$.  This implies
termination since infinite descent in a well-founded order is
impossible.
Unlike termination of programs in general, which is the fundamental example of undecidability, the algorithmic problems of detection
(deciding the existence) or generation (synthesis) of a ranking function can well be solvable, given certain choices of the program
representation, and the class of ranking function.  Numerous researchers have proposed such classes, with an eye towards decidability;
in some cases the algorithmic problems have been completely settled, and efficient algorithms provided, while other cases remain as
open problems.  Thus, in designing ranking functions, we look for expressivity (to capture more program behaviors) but also want
(efficient) computability.  Besides proving termination, some classes of ranking functions also serve to bound the length of the
computation (an \emph{iteration bound}), which is useful in applications such as \emph{cost analysis} (related terms:
execution-time analysis, resource analysis) and loop optimization \cite{Feautrier92.1,ADFG:2010,DBLP:journals/jar/AlbertAGP11,BrockschmidtE0F16}.

We focus on \emph{single-path linear-constraint loops} (\slc loops for short), where a
state is described by the values of a finite set of numerical
variables, and the effect of a transition (one iteration of the loop) is described by a conjunction of 
\emph{linear constraints}.
We consider the setting of integer-valued variables, as
well as rational-valued (or real-valued) variables%
\footnote{For the results in this paper, the real-number case is equivalent to the rational-number case, and in the sequel we
refer just to rationals.}.
Here is an example of this loop representation (a formal definition is in Section~\ref{sec:prelim});
primed variables $x',y',\dots$ refer to the state following the transition.
\begin{equation}
\label{eq:loop-xyz}
\while~(x \ge -z)~\wdo~x'=x+y,\; y'=y+z,\; z'=z-1
\end{equation}
Note that by $x'=x+y$ we mean an equation, not an assignment statement; it is a standard procedure to compile sequential code into such
equations (if the operations used are linear), or to approximate it using various techniques.

This constraint representation may be extended to represent branching in the loop body, a so-called \emph{multiple-path loop};
in the current work we do not consider such loops. However, \slc loops are important already, in particular in 
approaches that reduce a question about a whole program to questions about simple loops
\cite{Harrison1969book,leroux2005flat,CPR06,CGPRV2007,cookGLRS2008}; see \cite{Ouaknine2015siglog} for references that show the importance of
+such loops in other fields.
We assume the ``constraint loop" to be given, and do not concern ourselves with the orthogonal topic of extracting a loop representation from an actual program
(note that in some applications, such as analyzing dynamical systems of various kinds, we may start not with a computer program but with a model,
expressed by its creator as a set of constraints).

\paragraph*{Types of ranking functions.}
Several types of ranking
functions have been suggested; linear ranking functions (LRFs) are probably the most widely used and well-understood.
 In this case, we seek a function
$f(x_1,\dots,x_n) = a_1x_1+\dots+a_n x_n + a_0$, with the rationals
as a co-domain, such that 
\begin{inparaenum}[(i)]
\item\label{intro:lrf1} $f(\bar{x}) \ge 0$ for any valuation $\bar{x}$ that satisfies
  the loop constraints (i.e.,  an enabled state); and
\item\label{intro:lrf2} $f(\bar{x})-f(\bar{x}') \ge 1$ for any transition leading from $\bar{x}$ to
  $\bar{x}'$.
\end{inparaenum}
Technically, the rationals are not a well-founded set under the usual order, but we can refer to the partial order
$a \succeq b$ if and only if $a\ge 0$ and $a\ge b+1$, which is well-founded.
Given a linear-constraint loop, 
it is possible to find a linear ranking function (if one exists)
using linear programming (\lp).  This method
was found by multiple researchers in different places and times and
in some alternative versions~\cite{Feautrier92.1,DBLP:conf/pods/SohnG91,DBLP:conf/tacas/ColonS01,DBLP:conf/vmcai/PodelskiR04}.
Since \lp has a polynomial-time
complexity, most of these methods yield polynomial-time algorithms. This method is sound (any ranking function produced is valid),
and complete (if there is a ranking function, it will find one), when variables are assumed to range over the rationals. 
When variables range over the integers, treating the domain as $\rats$ is safe, but completeness is not guaranteed.
Consider the following loop:
\begin{equation}
\label{eq:intro:ints}
\while~( x_2-x_1 \le 0,\, x_1+x_2 \ge 1 ) ~\wdo~ x_2' = x_2-2x_1+1,\, x_1'=x_1
\end{equation}
and observe that it does not terminate over the rationals at all (try $x_1=x_2=\frac{1}{2}$); but it has
a \lrf that is valid for all integer valuations, e.g., $f(x_1,x_2) = x_1+x_2$.
Several authors noted this issue, and finally the
complexity of a complete solution for the integers was settled by~\cite{Ben-AmramG13jv},
who proved that the detection problem is \emph{{\coNP}-complete} and gave matching algorithms.

However, not all terminating loops
have a LRF; 
and to handle more loops, one may resort to an argument that combines several LRFs to capture
a more complex behavior. Two types of such behavior that re-occur in the literature on termination are
lexicographic ranking and multiphase ranking.

\textit{Lexicographic ranking}.  One can prove the termination of a loop by considering a tuple, say a pair
$\tuple{f_1,f_2}$ of linear functions, such that either $f_1$ decreases, or $f_1$ does not change and $f_2$ decreases.
There are some variants of the definition~\cite{DBLP:conf/cav/BradleyMS05,ADFG:2010,Ben-AmramG13jv,LarrazORR13}
regarding whether both functions have to be non-negative at
all times, or ``just when necessary."  
The most permissive definition allows any component to be negative, and technically, it ranks states in the lexicographic extension
of the order $\succeq$ mentioned above.  We refer to this class as \bmsllrfs, and are only aware of one work where this class is used in its full generatlity~\cite{LarrazORR13}.
For example, the following loop
\begin{equation}
\label{eq:loop-bms}
\while~(x \ge 0, y \le 10, z \ge 0, z\le 1)~\wdo~x'=x+y+z-10, y'=y+z, z'=1-z
\end{equation}
has the \llrf $\tuple{4y,4x-4z+1}$, which is  valid only according to the defintion of~\cite{LarrazORR13}, since it allows the first component to be negative for transitions that are ranked by the second component.

\textit{Multiphase ranking}. Consider loop~\eqref{eq:loop-xyz} above.
Clearly, the loop goes through three phases --- in the first, $z$ descends, while the other variables may increase; in the second (which begins once $z$ becomes
negative), $y$ decreases;
in the last phase (beginning when $y$ becomes negative), $x$ decreases.
 Note that since there is no lower bound on $y$ or on $z$, they cannot be used in a \lrf; however, 
each phase is clearly finite, as it is associated with a value that is non-negative and decreasing during that phase.
In other words, each phase is linearly ranked.
We shall say that this loop has the \emph{multiphase ranking function} (\mlrf) $\tuple{z+1,y+1,x}$.  The general definition
(Section~\ref{sec:prelim}) allows for an arbitrary number $d$ of linear components; we refer to $d$ as \emph{depth}, intuitively it is the number
of phases.

Some loops have multiphase behavior which is not so evident as in the last example. Consider
the following loop, that we will discuss further in Section~\ref{sec:depth}, with \mlrf $\tuple{x-4y , x-2y, x-y}$
\begin{equation}
\while~(x\ge 1,\; y\ge 1,\; x\ge y,\; 4 y\ge  x)~\wdo~x'=2x,\; y'=3y
\end{equation}
Technically, under which ordering is a \mlrf a ranking function? It is quite easy to see that the pairs used in the 
examples above descend in the lexicographic extension of $\succeq$. This means that \mlrfs are a sub-class of \bmsllrfs.
Note that, intuitively, a lexicographic ranking function also has ``phases", namely, steps where the first component decreases,
steps where the second component decreases, etc.; but these phases may alternate an unbounded number of times.

\paragraph*{Complete solutions and complexity.}
Complete solutions for \mlrfs (over the rationals) appear in~\cite{LeikeHeizmann15,li2016depth}. Both use non-linear constraint
solving, and therefore do not achieve a polynomial time complexity. \cite{BagnaraM13PPDP} study ``eventual linear ranking functions,"
which are \mlrfs of depth 2, and pose the question of a polynomial-time solution as an open problem, as well as the problem of a complete
solution for the integers.

In this paper, we provide complete solutions to the existence and synthesis problems for both 
\mlrfs and \llrfs, for  rational and integer \slc loops, 
where the algorithm is parameterized by a depth bound.  Over the rationals, the decision problem is PTIME and the synthesis can be done in polynomial
time; over the integers,  the existence problem is \coNP-complete, and our synthesis
procedure is deterministic exponential-time.

While such algorithms would be a contribution in itself, we find it even more interesting that our results are mostly based on discovering 
unexpected \emph{equivalences} between classes of ranking functions.  We prove two such results: Theorem~\ref{thm:wbms2mlrfQ} in Section~\ref{sec:bms-mlrf}
shows that \bmsllrfs are not stronger than \mlrfs for
\slc loops.  Thus, the complete solution for \bmsllrfs is just to solve for \mlrfs 
(for the loop \eqref{eq:loop-bms},
we find the \mlrf $\tuple{4y+x-z,4x-4z+4}$).
Theorem~\ref{thm:slc-nested} in Section~\ref{sec:ratcase} shows that one can further reduce the search for \mlrfs to a proper sub-class, called
\emph{nested} \mlrfs. This class was introduced in~\cite{LeikeHeizmann15} because its definition is simpler and allows for a polynomial-time
solution (over $\rats$). Thus, our equivalence result immediately implies a polynomial-time solution for \mlrfs.

Our complete solution for the \emph{integers} is also a reduction---transforming the problem so that solving over the rationals cannot give 
false alarms.  The transformation consists of computing the \emph{integer hull} of the transition polyhedron.  This transformation is 
well-known in the case of \lrfs \cite{Feautrier92.1,CookKRW10,Ben-AmramG13jv}, so it was a natural approach to try, however its proof in the case of \mlrfs is more involved.

We also make a contribution towards the use of \mlrfs in deriving \emph{iteration bounds}.  
As the loop \eqref{eq:loop-xyz}
demonstrates, it is possible for the variables that control subsequent phases to grow (at a polynomial rate)  during the first phase.
Nonetheless, we prove that \emph{any \mlrf implies a linear bound on the number of iterations} for a \slc loop (in terms of the initial
values of the variables).  Thus, it is also the case that any \bmsllrf implies a linear bound.

An open problem raised by our work is whether one can precompute a bound on the depth of a \mlrf for a given loop (if there is one);
for example~\cite{Ben-AmramG13jv} prove a depth bound of $n$ (the number of variables) on their notion of \llrfs (which is more restrictive); however their class is known to be weaker
than \mlrfs and \bmsllrfs. In Section~\ref{sec:depth} we discuss this problem.

The rest of this article is organized as follows.
Section~\ref{sec:prelim} gives precise definitions and some technical background necessary for this work. Sections~\ref{sec:ratcase} and~\ref{sec:bms-mlrf}
give our equivalence results for different types of ranking functions (over the rationals) and the algorithmic implications.
Section~\ref{sec:intcase} covers the integer setting, Section~\ref{sec:depth} discusses depth bounds, Section~\ref{sec:iter-bound} discusses the iteration bound, and Section~\ref{sec:conc} concludes.

%% file: preliminaries.tex

\section{Preliminaries}
\label{sec:prelim}

In this section we 
give the fundamental definitions for this paper: we define the class of loops
we study, the type of ranking functions, and
recall some definitions regarding (integer) polyhedra.
We also mention some important properties of these definitions.

\subsection{Single-Path Linear-Constraint Loops}
\label{sec:prelim:slcloops}

A \emph{single-path} linear-constraint loop (\slc for short) over
$n$ variables $x_1,\ldots,x_n$ has the form
\begin{equation} \label{eq:ilc-loop} 
  \mathit{while}~(B\vec{x} \le
  \vec{b})~\mathit{do}~ A\begin{pmatrix}\vec{x}\phantom{'}\\
    \vec{x}'\end{pmatrix} \le \vec{c}
\end{equation}
where $\vec{x}=(x_1,\ldots,x_n)^\trans$ and
$\vec{x}'=(x_1',\ldots,x_n')^\trans$ are column vectors, and for some
$p,q>0$, $B \in {\rats}^{p\times n}$, $A\in {\rats}^{q\times 2n}$,
$\vec{b}\in {\rats}^p$, $\vec{c}\in {\rats}^q$.
The constraint $B\vec{x} \le \vec{b}$ is called \emph{the loop
  condition} (a.k.a. the loop guard) and the other constraint is
called \emph{the update}. 
We say that the loop is a \emph{rational loop} if $\vec{x}$ and
$\vec{x}'$ range over $\rats^n$, and that it is an \emph{integer loop}
if they range over $\ints^n$.  One could also allow variables to take any real-number values,
but as long as the constraints are expressed by rational numbers this makes no difference
from the rational case. 

We say that there is a transition from a state $\vec{x}\in\rats^n$ to
a state $\vec{x}'\in\rats^n$, if $\vec{x}$ satisfies the condition and
$\vec{x}$ and $\vec{x}'$ satisfy the update.
A transition can be seen as a point $\trcv{\vec{x}}{\vec{x}'} \in \rats^{2n}$, where its first $n$
components correspond to $\vec{x}$ and its last $n$ components to
$\vec{x}'$. For ease of notation, we denote $\tr{\vec{x}}{\vec{x}'}$ by
$\vec{x}''$.
The set of all transitions $\vec{x}''\in \rats^{2n}$, of a given \slc
loop, will be denoted by $\transitions$ and is specified by the set
of inequalities $A'' \vec{x}'' \le \vec{c}''$ where
\begin{align*}
A''  & = \begin{pmatrix} B & 0 \\ \multicolumn{2}{c}{A} \end{pmatrix}  &
\vec c'' & = \begin{pmatrix} \vec b \\ \vec c \end{pmatrix}
\end{align*}
We call $\transitions$ \emph{the transition polyhedron} (for definitions regarding polyhedra see Sect.~\ref{sec:prelim:poly}).
For the purpose of this article, the essence of the loop is this polyhedron, even if the loop is
presented in a more readable form as \eqref{eq:ilc-loop}.

\subsection{Multi-Phase Ranking Functions}

An affine function $f: \rats^n \to \rats$ is of the form
$f(\vec{x}) = \vect{\rfcoeff}\cdot\vec{x} + \rfcoeff_0$ where
$\vect{\rfcoeff}\in\rats^n$ is a row vector and $\rfcoeff_0\in\rats$.
For a given function $f$, we define the function
$\diff{f}:\rats^{2n}\mapsto\rats$ as
$\diff{f}(\vec{x}'')=f(\vec{x})-f(\vec{x}')$.

\bdfn[\mlrf]
\label{def:mlrf}
Given a set of transitions $T\subseteq\rats^{2n}$, we say that
$\mlrfsym=\tuple{f_1,\dots,f_d}$ is a \mlrf (of depth $d$) for $T$ if for
every $\vec{x}'' \in T$ there is an index $i\in[1,d]$ such that:
\begin{alignat}{ 2 }
 \forall j \leq i \ .\   && \diff{f_j}(\vec{x}'') & \geq 1 \,, \label{eq:mlrf:1}\\[-0.5ex]
                      && f_i(\vec{x}) &\ge 0          \,, \label{eq:mlrf:2}\\[-0.5ex]
  \forall j < i \ .\                    && f_j(\vec{x}) &\le 0\,. \label{eq:mlrf:3} 
\end{alignat}
We say that $\vec{x}''$ is \emph{ranked by} $f_i$ (for the minimal such $i$).
\edfn

It is not hard to see that this definition, for $d=1$, means that $f_1$ is a linear ranking function, and
for $d>1$, it implies that 
as long as $f_1(\vec x) \ge 0$, transition $\vec{x}''$ must be ranked by $f_1$, and when $f_1(\vec x) < 0$,
$\tuple{f_2,\dots,f_d}$ becomes a \mlrf. This agrees with the intuitive notion of a ``phases."
We further note that, for loops specifies by polyhedra, making the inequality \eqref{eq:mlrf:3} strict results in the same class of ranking
functions (we chose the definition that is easier to work with), and, similarly,
we can  replace~\eqref{eq:mlrf:1} by
$\diff{f_j}(\vec{x}'') > 0$, obtaining an equivalent definition (up to multiplication of the $f_i$ by some constants).
We say that $\mlrfsym$ is \emph{irredundant} if removing any component
invalidates the \mlrf.
Finally, it is convenient to allow an empty tuple as a \mlrf for the empty set.

The decision problem \emph{Existence of a \mlrf} asks to determine
whether a given \slc loop admits a \mlrf. We denote this problem by
\mlrfq and \mlrfz for rational and integer loops respectively.
The \emph{bounded} decision problem, denoted by \dmlrfq and \dmlrfz,
restricts the search to \mlrfs of depth at most $d$, where the parameter $d$ is part of the input.

\subsection{Polyhedra}
\label{sec:prelim:poly}

A \emph{rational convex polyhedron} $\poly{P} \subseteq \rats^n$
(\emph{polyhedron} for short) is the set of solutions of a set of
inequalities $A\vec{x} \le \vec{b}$, namely $\poly{P}=\{
\vec{x}\in\rats^n \mid A\vec x \le \vec b \}$, where $A \in \rats^{m
  \times n}$ is a rational matrix of $n$ columns and $m$ rows, $\vec
x\in\rats^n$ and $\vec b \in \rats^m$ are column vectors of $n$ and
$m$ rational values respectively.
We say that $\poly{P}$ is specified by $A\vec{x} \le \vec{b}$.
If $\vec{b}=\cvz$, then $\poly{P}$ is a \emph{cone}.
The set of \emph{recession directions} of a polyhedron $\poly{P}$
specified by $A\vec{x} \le \vec b$, also know as its \emph{recession
  cone}, is the set $\ccone(\poly{P}) = \{ \vec{y}\in\rats^n \mid
A\vec{y} \le \vec{0}\}$.

For a given polyhedron $\poly{P} \subseteq \rats^n$ we let
$\intpoly{\poly{P}}$ be $\poly{P} \cap \ints^n$, i.e., the set of
integer points of $\poly{P}$. The \emph{integer hull} of $\poly{P}$,
commonly denoted by $\inthull{\poly{P}}$, is defined as the convex
hull of $\intpoly{\poly{P}}$, i.e., every rational point of
$\inthull{\poly{P}}$ is a convex combination of integer points.
It is known that $\inthull{\poly{P}}$ is also a polyhedron, and that
$\ccone(\poly{P})=\ccone(\inthull{\poly{P}})$.
An \emph{integer polyhedron} is a polyhedron $\poly{P}$ such that
$\poly{P} = \inthull{\poly{P}}$. We also say that $\poly{P}$ is
\emph{integral}.

Polyhedra also have a \emph{generator representation} in terms of
vertices and rays, written as
$\poly{P} = \convhull\{\vec x_1,\dots,\vec x_m\} + \cone\{\vec
y_1,\dots,\vec y_t\} \,.$
This means that $\vec x\in \poly{P}$ iff $\vec x = \sum_{i=1}^m
a_i\cdot \vec x_i + \sum_{j=1}^t b_j\cdot \vec y_j$ for some rationals
$a_i,b_j\ge 0$, where $\sum_{i=1}^m a_i = 1$.  Note that $\vec
y_1,\dots,\vec y_t$ are the recession directions of $\poly{P}$, i.e.,
$\vec{y}\in\ccone(\poly{P})$ iff $\vec{y}=\sum_{j=1}^t b_j \cdot
\vec{y}_j$ for some rationals $b_j\ge 0$.
If $\poly{P}$ is integral, then there is a generator representation in
which all $\vec{x}_i$ and $\vec{y}_j$ are integer.

Next we state some lemmas that are fundamental for many proofs in this
article. 
Given a polyhedron $\poly{P}$, the lemmas show that if a disjunction of constraints of the form
$f_i > 0$, or $f_i \ge 0$, holds over $\poly{P}$, then a certain conic combination of these functions is 
positive (or non-negative) over $\poly{P}$.
 This generalizes Lemma~1
of~\cite{HeizmannHLP:ATVA2013}.
The lemmas are all very similar, but vary in the the use strict or non-strict
inequalities.

\blem
\label{lem:posfuncstr}
Given a non-empty polyhedron $\poly{P}$, and linear functions $f_1,\ldots,f_k$
such that
\begin{enumerate}[\upshape(\itshape i\upshape)]
\item \label{posfuncstrAss1}
 $\vec{x}\in\poly{P} \rightarrow f_1(\vec{x}) > 0
  \vee\cdots\vee f_{k-1}(\vec{x}) > 0 \vee f_k(\vec{x}) \geq 0$
\item $\vec{x}\in\poly{P} \not\rightarrow f_1(\vec{x}) >  0
  \vee\cdots\vee f_{k-1}(\vec{x}) > 0$
\end{enumerate}
There exist non-negative constants $\mu_1,\ldots,\mu_{k-1}$ such
that
$$\vec{x}\in\poly{P} \rightarrow \mu_1 f_1(\vec{x})+\cdots+
 \mu_{k-1} f_{k-1}(\vec{x})+ f_{k}(\vec{x}) \geq 0 \,.$$
\elem

\bprf
Let $\poly{P}$ be $B\vec{x} \le \vec{c}$,
$f_i = \vect{a}_i\cdot\vec{x}-b_i$, then (\textit{i}) is equivalent to infeasibility of
\begin{equation}
  B\vec{x} \le \vec{c} \wedge A\vec{x} \le \vec{b} \wedge \vect{a}_k\cdot\vec{x} < b_k
\end{equation}
where $A$ consists of the $k-1$ rows $\vect{a}_i$, and $\vec{b}$ of corresponding $b_i$.
However, $B\vec{x}\le \vec{c} \land A\vec{x}\le \vec{b}$ is assumed to be feasible.

According to Motzkin's transposition theorem~\cite[Corollary 7.1k,
Page~94]{schrijver86}, this implies that there are row vectors
$\vect{\lambda}, \vect{\lambda}' \ge 0$ and a constant $\mu \ge 0$
such that the following is true:
\begin{equation}
\label{eq:Motz1}
\vect{\lambda}B+\vect{\lambda}'A + \mu a_k = 0 \land \vect{\lambda}\vec{c} + \vect{\lambda}'\vec{b} + \mu b_k \le 0 
 \land  ( \mu\ne 0 \lor  \vect{\lambda}\vec{c} + \vect{\lambda}'\vec{b} + \mu{b_k} < 0 )
\end{equation}
Now, if \eqref{eq:Motz1} is true, then for all $\vec{x}\in\poly{P}$,
\begin{align*}
(\sum_i \lambda'_i f_i (\vec{x}) ) + \mu f_k(\vec{x}) &= \vect{\lambda}'A\vec{x} - \vect{\lambda}'\vec{b}  + \mu a_k\vec{x} - \mu b_k \\
& = -\vect{\lambda}B\vec{x} - \vect{\lambda}'\vec{b} -\mu b_k  \ge  -\vect{\lambda}\vec{c} - \vect{\lambda}' \vec{b}  - \mu b_k \ge 0
\end{align*}
where if $\mu = 0$, the last inequality must be strict.  However, if
$\mu = 0$, then $\vect{\lambda}B + \vect{\lambda}'A = 0$, so by
feasibility of $B\vec{x}\le \vec{c}$ and $A\vec{x}\le \vec{b}$, 
this implies  
 $\vect{\lambda}\vec{c} + \vect{\lambda}'\vec{b} \ge 0$, a
contradiction.  Thus, $(\sum_i \lambda'_i f_i ) + \mu f_k\ge 0$ on
$\poly{P}$ and $\mu> 0$. Dividing by $\mu$ we obtain the conclusion of
the lemma.  
\eprf

\blem
\label{lem:posfunc}
Given a non-empty polyhedron $\poly{P}$, and linear functions $f_1,\ldots,f_k$
such that
\begin{enumerate}[\upshape(\itshape i\upshape)]
\item
 $\vec{x}\in\poly{P} \rightarrow f_1(\vec{x}) \geq 0
  \vee\cdots\vee f_k(\vec{x}) \geq 0$
\item $\vec{x}\in\poly{P} \not\rightarrow f_1(\vec{x}) \geq 0
  \vee\cdots\vee f_{k-1}(\vec{x}) \geq 0$
\end{enumerate}
There exists non-negative constants $\mu_1,\ldots,\mu_{k-1}$ such
that
$$\vec{x}\in\poly{P} \rightarrow \mu_1 f_1(\vec{x})+\cdots+
 \mu_{k-1} f_{k-1}(\vec{x})+ f_{k}(\vec{x}) \geq 0 \,.$$
\elem

\bprf
Let $\poly{P}$ be $B\vec{x} \le \vec{c}$,
$f_i = \vect{a}_i\vec{x}-b_i$, then (\textit{i}) is equivalent to infeasibility of
%
\(
  B\vec{x} \le \vec{c} \wedge A\vec{x} < \vec{b}
\).
According to Motzkin's transposition theorem, this implies that there are row vectors
$\vect{\lambda}, \vect{\mu} \ge 0$ such that the following is true:
\begin{equation}
\label{eq:Motz2}
\vect{\lambda}B+\vect{\mu}A = 0 \land \vect{\lambda}\vec{c} + \vect{\mu}\vec{b} \le 0 \quad
 \land  ( \vect{\mu}\ne 0 \lor  \vect{\lambda}\vec{c} + \vect{\mu}\vec{b} < 0 )
\end{equation}
Now, if \eqref{eq:Motz2} is true, then for all $\vec{x}\in\poly{P}$,
\[
\sum_i \mu_i f_i(\vec{x})  = \vect{\mu}A\vec{x} - \vect{\mu}\vec{b} = -\vect{\lambda}B\vec{x} - \vect{\mu}\vec{b}  \ge -\vect{\lambda}\vec{c} - \vect{\mu} \vec{b} \ge 0
\]
where if $\vect{\mu}=0$, the last inequality must be strict.
However, if $\vect{\mu}=0$, then $\vect{\lambda}B = 0$, so by feasibility of $\poly{P}$ and 
feasibility of $B\vec{x}\le \vec{c}$ and $A\vec{x}\le \vec{b}$, 
this implies  
$\vect{\lambda}\vec{c} \ge 0$, a contradiction.
Thus, $\sum_i \mu_i f_i \ge 0$ on $\poly{P}$ and $\vect{\mu}\ne 0$. Based on assumption (\textit{ii}),
such a combination must include $f_k$ with a positive coefficient, and therefore can be normalized to the stated form.
\eprf

%% file: nested.tex

\section{Complexity of  Synthesis of \mlrfs over the Rationals}
\label{sec:ratcase}

In this section we study the complexity of deciding if a given rational \slc
loop has a \mlrf of depth $d$, and show that this  can
be done in polynomial time.
These results follow from an equivalence between \mlrfs and
a sub-class called \emph{nested ranking functions}~\cite{LeikeHeizmann15}.
In the rest of this article we assume a given \slc loop specified by a
transition polyhedron $\transitions$. The complexity results assume a constraint representation for $\transitions$, as in Section~\ref{sec:prelim:slcloops}.

\bdfn
\label{def:nested-mlrf}
A $d$-tuple $\llrfsym = \tuple{f_1,\dots,f_d}$ is a \emph{nested
  ranking function} for $\transitions$ if the following requirements
are satisfied for all $\vec{x}'' \in \transitions$
\begin{alignat}{2}
f_d(\vec{x}) \ge 0\label{eq:lastPositive} & &\\
(\diff{f_i}(\vec{x}'') - 1) + f_{i-1}(\vec{x}) \ge 0  &\quad\quad\quad& \mbox{for all } i=1,\dots,d.\label{eq:nested}
\end{alignat}
where for uniformity we define $f_0 \equiv 0$.
\edfn

It is easy to see that a nested ranking function is a
\mlrf. Indeed, $f_1$ is decreasing, and when it becomes negative
$f_2$ starts to decrease, etc. In addition, the loop must stop by  the time that
the last component becomes negative, since $f_d$ is non-negative on all
enabled states.

\bexm
\label{ex:loop1}
Consider loop~\eqref{eq:loop-xyz} (on Page~\pageref{eq:loop-xyz}).
It has the \mlrf $\tuple{z+1,y+1,x}$ which is not nested because, among
other things, last component $x$ might be negative, e.g., for the
state $x=-1, y=0, z=1$.
However, it has the nested ranking function $\tuple{z+1,y+1,z+x}$,
which is \mlrf.
\eexm

The above example shows that there are \mlrfs which are not nested
ranking functions, however, next we show that if a loop has a \mlrf
then it has (possibly different) nested ranking function of the same depth. We first
state an auxiliary lemma, and then prove the main result.

\blem
\label{lem:mlrf-pos}
Let $\llrfsym=\tuple{f_1,\ldots,f_d}$ be an irredundant \mlrf for
$\transitions$, such that
$\tuple{f_2,\dots,f_d}$ is a nested ranking function for
$\transitions' = \transitions \cap \{\vec{x}''\in \rats^{2n} \mid f_1(\vec
x) \le 0 \}$.
Then there is a nested ranking function of depth $d$ for
$\transitions$.
\elem

\bprf
First recall that, by definition of \mlrf, we have $\diff{f_1}(\vec{x}'')\ge 1$
for any $\vec{x}''\in\transitions$, and since $\tuple{f_2,\dots,f_d}$
is a nested ranking function for $\transitions'$ we have
\begin{equation}
\label{eq:ih:nested}
\begin{split}
\vec{x}''\in \transitions' \rightarrow 
 & f_d(\vec{x}) \ge 0\\
\vec{x}''\in \transitions' \rightarrow 
  & (\diff{f_2}(\vec{x}'')-1)+f_0(\vec{x}'') \ge 0 \ \wedge\\
  & (\diff{f_3}(\vec{x}'')-1)+f_2(\vec{x}'') \ge 0 \ \wedge\\
  & \vdots\\
  & (\diff{f_d}(\vec{x}'')-1)+f_{d-1}(\vec{x}'') \ge 0
\end{split}
\end{equation}
Next we construct a nested ranking function $\tuple{f_1',\ldots,f_d'}$
for $\transitions$, i.e., such that~\eqref{eq:lastPositive} is
satisfied for $f_d'$, and~(\ref{eq:nested}) is satisfied for each
$f'_i$ and $f'_{i-1}$ --- we refer to the instance of (\ref{eq:nested}) for a specific
$i$ as (\ref{eq:nested}$_{i}$).

We start with the condition~\eqref{eq:lastPositive}. If $f_d$ is
non-negative over $\transitions$ we let $f_d'=f_d$, otherwise, clearly
$$\vec{x}''\in \transitions \rightarrow f_d(\vec x) \ge 0 \lor  f_1(\vec x) > 0 \,.$$
Then, by Lemma~\ref{lem:posfuncstr} there is a constant $\mu_d>0$ such that 
$$\vec{x}''\in \transitions \rightarrow f_d(\vec x)+ \mu_d f_1(\vec x) \ge 0$$
and we define $f_d'(\vec x) = f_d(\vec x) + \mu_d f_1(\vec
x)$. Clearly~\eqref{eq:lastPositive} holds for $f_d'$.

Next, we handle the conditions~(\ref{eq:nested}$_{i}$) for $i=d,\dots,
3$ in this order. When we handle condition~(\ref{eq:nested}$_{i}$), we
shall define $f'_{i-1}(\vec x) = f_{i-1}(\vec x)+\mu_{i-1} f_1(\vec x)$ for
some $\mu_{i-1} \ge 0$. Note that $f'_d$ has this form as well.

Suppose we have computed $f_d',\dots,f_{i}'$. We show how to ensure that 
(\ref{eq:nested}$_{i}$) holds over $\transitions$ by
constructing $f'_{i-1}$.
From~\eqref{eq:ih:nested} we know that 
$$\vec{x}''\in \transitions' \rightarrow (\diff{f_i}(\vec{x}'')-1)+f_{i-1}(\vec{x}'') \ge 0 \,.$$
Now since $f'_{i}(\vec x) = f_i(\vec x)+\mu_i f_1(\vec x)$, and
$\diff{f_1}(\vec{x}'') \ge 1$ over $\transitions$, we have 
$$\vec{x}''\in \transitions' \rightarrow
(\diff{f'_i}(\vec{x}'')-1)+f_{i-1}(\vec{x}'') \ge 0 \,.$$
Now if ($\diff{f'_i}(\vec{x}'')-1)+f_{i-1}(\vec{x}'') \ge 0$ holds
over $\transitions$ as well, we let $f_{i-1}'=f_{i-1}$. Otherwise, we
have
$$\vec{x}''\in\transitions \rightarrow (\diff{f'_i}(\vec{x}'') - 1) + f_{i-1}(\vec x)  \ge 0 \lor f_1(\vec{x})>0 \,,$$
and by Lemma~\ref{lem:posfuncstr} there is $\mu_{i-1} > 0$ such that
$$\vec{x}''\in\transitions \rightarrow (\diff{f'_i}(\vec{x}'') - 1) + f_{i-1}(\vec x) + \mu_{i-1} f_1(\vec x) \ge 0 \,.$$
In this case, we let $f'_{i-1}(\vec x) = f_{i-1}(\vec x) + \mu_{i-1} f_1(\vec x)$.
Clearly~(\ref{eq:nested}$_{i}$) holds.

Next we proceed to~(\ref{eq:nested}$_{2}$). From~\eqref{eq:ih:nested}
we know that
$$\vec{x}''\in \transitions' \rightarrow (\diff{f_2}(\vec{x}'')-1)+f_{1}(\vec{x}'') \ge 0 \,.$$
Since $f_2'=f_2+\mu_2 f_1$ and $\diff{f_1}(\vec{x}'') \ge 1$
we have
$$\vec{x}''\in\transitions' \rightarrow (\diff{f'_2}(\vec{x}'') - 1) + f_{1}(\vec{x}'') \ge 0 \,.$$
Next, by definition of $\transitions'$ and the lemma's assumption we have
$$\vec{x}''\in\transitions \rightarrow (\diff{f'_2}(\vec{x}'') - 1) \ge 0 \lor f_1(\vec{x}) > 0$$
and we also know that
$(\diff{f'_2}(\vec{x}'') - 1) \ge 0$ does not hold over
$\transitions$, because then $f_1$ would be redundant. Now by Lemma~\ref{lem:posfuncstr} there is $\mu_1 >
0$ such that
$$\vec{x}''\in\transitions \rightarrow (\diff{f'_2}(\vec{x}'') - 1) + \mu_1 f_1(\vec x) \ge 0 \,.$$
We let $f_1'(\vec x)=\mu_1 f_1(\vec x)$. 
For~(\ref{eq:nested}$_{1}$) we need to show that $\diff{f'_1}(\vec{x}'')
- 1 \ge 0$ holds over $\transitions$, which clearly true if $\mu_1\ge
1$ since $\diff{f}_1\ge 1$, otherwise we multiply all $f_i'$ by
$\frac{1}{\mu_1}$ which does not affect any~(\ref{eq:nested}$_{i}$)
and makes~(\ref{eq:nested}$_{1}$) true.
\eprf

\bthm 
\label{thm:slc-nested}
If $\transitions$ has a \mlrf of depth $d$, then it has a nested ranking function of depth at most $d$.
\ethm

\bprf
The proof is by induction on $d$.  We assume a \mlrf
$\tuple{f_1,\ldots,f_d}$ for $\transitions$.
For $d=1$ there is no difference between a general \mlrf and a nested
one.  
For $d>1$, we consider $\tuple{f_2,\dots,f_d}$ as a \mlrf for
$\transitions' = \transitions \cap \{ \vec{x}''\in\rats^{2n} \mid
f_1(\vec x) \le 0\}$, we apply the induction hypothesis to turn
$\tuple{f_2,\dots,f_d}$ into a nested ranking function. Either $f_1$ becomes redundant, or we can apply
Lemma~\ref{lem:mlrf-pos}.
\eprf

The above theorem give us a complete algorithm for the synthesis of
\mlrfs of a given depth $d$ for $\transitions$, i.e., for rational \slc
loops, namely, we synthesize a nested ranking function. 

\bthm
$\dmlrfq\in\ptime$.
\ethm

\bprf
We describe, in some detail, how to synthesize a nested ranking function in polynomial time
(this actually appears in~\cite{LeikeHeizmann15}). Due to Theorem~\ref{thm:slc-nested}, this yields a complete decision procedure for \mlrfs.
Given $\transitions$, our goal is to find  $f_1,\ldots,f_d$ such
that~(\ref{eq:lastPositive},\ref{eq:nested}) hold.
If we take just one of the conjuncts, our task is to find coefficients for the functions involved ($f_d$, or $f_i$ and $f_{i-1}$),
such that the desired inequality is implied by $\transitions$. Using Farkas' lemma~\cite{schrijver86},
this problem can be formulated as a \lp problem,
where the coefficients we seek are unknowns. By conjoing all these \lp problems, we obtain a single \lp problem, 
of polynomial size, whose solution---if there is one---provides the coefficients of all $f_i$; and if there is no solution,
then no nested ranking function exists. Since \lp is polynomial-time, this procedure has
polynomial time complexity.
\eprf

Clearly, if $d$ is considered as
constant, then $\dmlrfq$ is polynomial in the bit-size of the input
$\transitions$.
When considering $d$ as variable, then the complexity is polynomial in the
bit-size of $\transitions$ plus $d$---equivalently, it 
is polynomial in the bit-size of the
input if we assume that $d$ is given
in unary representation (which is a reasonable assumption since $d$ describes the number of components of
the  \mlrf sought). The same observation applies to our classification of $\dmlrfz$ (Section~\ref{sec:intcase}).

%% file: mlrf-vs-bms.tex

\section{Multiphase vs Lexicographic-Linear Ranking Functions}
\label{sec:bms-mlrf}

\mlrfs are similar to \bmsllrfs, and a natural question is: which one
is more powerful for proving termination of \slc loops? In this
section we show that they have the same power, by proving that an \slc
has a \mlrf if and only if it has a \bmsllrf.
We first note that there are several definitions for
\bmsllrfs~\cite{DBLP:conf/cav/BradleyMS05,ADFG:2010,Ben-AmramG13jv,LarrazORR13}.
The following is the most general one~\cite{LarrazORR13}.

\bdfn
\label{def:llrf}
Given a set of transitions $T\subseteq\rats^{2n}$, we say that
$\tuple{f_1,\dots,f_d}$ is a \bmsllrf (of depth $d$) for $T$ if for
every $\vec{x}'' \in T$ there is an index $i$ such that:
\begin{alignat}{ 2 }
 \forall j < i \ .\   && \diff{f_j}(\vec{x}'') & \geq 0 \,, \label{eq:llrf:1}\\[-0.5ex]
                      && \diff{f_i}(\vec{x}'') & \geq 1 \,, \label{eq:llrf:2}\\[-0.5ex]
                      && f_i(\vec{x}) &\ge 0          \,, \label{eq:llrf:3}
\end{alignat}
We say that $\vec{x}''$ is \emph{ranked by} $f_i$ (for the minimal
such $i$).  
\edfn

Regarding other definitions: \cite{Ben-AmramG13jv}
requires~\eqref{eq:llrf:3} to hold for all $f_j$ with $j\leq i$, and~\cite{ADFG:2010}
requires~\eqref{eq:llrf:3} to hold for all
components. They are clearly more restrictive.
Actually~\cite{ADFG:2010} shows that an \slc loop has a \bmsllrf
according to their definition if and only if it has a \lrf, which is not the
case of~\cite{Ben-AmramG13jv}.
The definition of~\cite{DBLP:conf/cav/BradleyMS05} is equivalent to
a \lrf when considering \slc loops, as their main interest is in
multipath loops.

It is easy to see that a \mlrf is also a \llrf as in
Definition~\ref{def:llrf}. Next we show that for \slc loops any
\bmsllrf can be converted to a \mlrf, proving that these classes of
ranking functions have the same power for \slc loops. We start with an
auxiliary lemma.

\blem
\label{lem:equiv:1}
Let $f$ be a non-negative linear function over $\transitions$.  If
$\transitions'  =  \transitions\cap \{\vec{x}''\mid{\diff{f}(\vec{x}'')\leq 0}\}$ has a
\mlrf of depth $d$, then $\transitions$ has a \mlrf of depth at most $d+1$. 
\elem

\bprf
Note that simply appending $f$ to a \mlrf $\tau$ of $\transitions'$
does not always produce a \mlrf, since the components of $\tau$ are
not guaranteed to decrease over $\transitions\setminus\transitions'$.
Let $\tau=\tuple{g_1,\ldots,g_d}$ be a \mlrf for $\transitions'$, we
show how to construct a \mlrf $\tuple{g_1',\ldots,g_d',f}$.
If $g_1$ is decreasing over $\transitions$, we define
$g_1'(\vec{x})=g_1(\vec{x})$, otherwise we have
$$\vec{x}''\in \transitions \rightarrow \diff{f}(\vec{x}'')>0
\vee \diff{g_1}(\vec{x}'') \geq 1,$$ 
then by Lemma~\ref{lem:posfuncstr} we can construct $g_1'(\vec{x})=g_1(\vec{x})+\mu f(\vec{x})$ such that
$\vec{x}''\in \transitions \rightarrow \diff{g_1'}(\vec{x}'') \geq 1$. Moreover,
since $f$ is non-negative $g_1'$ is non-negative on the transitions on
which $g_1$ is non-negative.
If $d>1$, we proceed with $\transitions^{(1)} = \transitions \cap
\{\vec{x}''\mid g_1'(\vec{x}) \le (-1)\}$.  Note that these transitions must be ranked,
in $\transitions'$, by $\tuple{g_2,\dots,g_d}$.  
If $g_2$ is decreasing over $\transitions^{(1)}$, let $g_2'=g_2$,
otherwise 
$$\vec{x}''\in \transitions^{(1)} \rightarrow \diff{f}(\vec{x}'')>0
\vee \diff{g_2}(\vec{x}'') \geq 1,$$
and again by Lemma~\ref{lem:posfuncstr} we can construct the desired
$g_2'$.
In general for any $j\le d$ we construct $g_{j+1}'$ such that
$\diff{g_{j+1}'} \ge 1$ over
$$\transitions^{(j)} = \transitions  \cap \{\vec{x}''\in\rats^{2n}\mid g_1'(\vec{x}) \le (-1) \wedge \dots \wedge g_{j}'(\vec{x}) \le (-1) \}$$
and $\vec{x}''\in \transitions \land g_j(\vec{x}) \ge 0 \rightarrow g_j'(\vec{x}) \ge 0$.
Finally we define
$$\transitions^{(d)} = \transitions  \cap \{\vec{x}''\in\rats^{2n}\mid g_1'(\vec{x}) \le (-1) \wedge g_d'(\vec{x}) \le (-1) \},$$
each transition $\vec{x}''\in\transitions^{(d)}$ must satisfy
$\diff{f}(\vec{x}'')>0$, and in such case $\diff{f}(\vec{x}'')$ must
have a minimum $c > 0$ since $\transitions^{(d)}$ is a polyhedron.
Without loss of generality we assume $c \ge 1$, otherwise take
$\frac{1}{c}f$ instead of $f$ .
Now $\tau'=\tuple{g_1'+1,\dots,g_d'+1,f}$ is a \mlrf for
$\transitions$. Note that if we arrive to $\transitions^{(j)}$ that is
empty, we can stop since we already have a \mlrf. 
\eprf

In what follows, by a \emph{weak} \bmsllrf we mean the class of
ranking functions obtained by changing condition~\eqref{eq:llrf:2} to
$\diff{f_i}(\vec{x}'') > 0$. Clearly it is more general than
\bmsllrfs, and as we will see next it suffices guarantees termination since we show how to convert them to \mlrfs.
 We prefer to use this class as it simplifies the proof of
the integer case that we present in Section~\ref{sec:intcase}.

\blem
\label{lem:equiv:2}
Let $\tuple{f_1,\ldots,f_d}$ be a \emph{weak} \bmsllrf for
$\transitions$. There is a linear function $g$ that is positive over
$\transitions$, and decreasing on (at least) the same transitions of
$f_i$, for some $1\le i\le d$.
\elem

\bprf
If any $f_i$ is positive over $\transitions$,
we take $g=f_i$. Otherwise,
we have $\vec{x}''\in \transitions \rightarrow f_1(\vec{x})\geq 0 \vee \cdots \vee
f_d(\vec{x}) \geq 0$ since every transition is ranked by some $f_i$.
If this implication satisfies the conditions of
Lemma~\ref{lem:posfunc} then we can construct $g(\vec{x})=f_d(\vec{x})+\sum_{i=1}^{d-1}\mu_if_i(\vec{x})$ that is non-negative
over $\transitions$, and, moreover, decreases on the
transitions that are ranked by $f_d$.
If the conditions of Lemma~\ref{lem:posfunc} are not satisfied, then the second condition
must be false, that is, $\vec{x}''\in \transitions \rightarrow f_1(\vec{x})\geq 0 \vee \cdots \vee
f_{d-1}(\vec{x})\geq 0$.  
Now we repeat the same reasoning as above for this implication. Eventually we
either construct $g$ that corresponds for some $f_i$ as above, or we arrive to
$\vec{x}''\in \transitions \rightarrow f_1(\vec{x})\geq 0$, and then take $g=f_1$.
\eprf

\bthm
\label{thm:wbms2mlrfQ}
If $\transitions$ has a weak \bmsllrf of depth $d$, 
it has a
\mlrf of depth $d$.
\ethm

\bprf
Let $\tuple{f_1,\ldots,f_d}$ be a weak
\bmsllrf for $\transitions$.    
We show how to
construct a corresponding \mlrf.

The proof is by induction on the depth $d$ of the \bmsllrf. For $d=1$
it is clear since it is an \lrf.
Now let $d>1$, by Lemma~\ref{lem:equiv:2} we can find $g$ that is
positive over $\transitions$ and decreasing at least on the same
transitions as $f_i$.
Now $\tuple{f_1,\ldots,f_{i-1},f_{i+1},f_d}$ is a weak \bmsllrf of
depth $d-1$ for $\transitions'= \transitions \cap \{\vec{x}''\mid
\diff{g}(\vec{x''})\le 0 \}$.
By the induction hypothesis we can construct a weak \mlrf for
$\transitions'$ of depth $d-1$, and then by Lemma~\ref{lem:equiv:1} we
can lift this \mlrf to one of depth $d$ for $\transitions$.
\eprf

\bexm
\label{def:bms2mlrf}
Let $\transitions$ be the transition polyhedron of the
loop~\eqref{eq:loop-bms} (on Page~\pageref{eq:loop-bms}),
which admits the \bmsllrf $\tuple{4y,4x-4z+4}$, and note that it is
not a \mlrf. Following the proof of above theorem,
we can convert it to the \mlrf $\tuple{4y+x-z,4x-4z+4}$.
\eexm

%% file: integer-mlrfs-bmsllrfs.tex

\section{\mlrfs and \bmsllrfs Over the Integers}
\label{sec:intcase}

The procedure described in Section~\ref{sec:ratcase} for synthesizing
\mlrfs, i.e., use linear programming to synthesize a nested ranking function, is
complete for rational loops but not for integer loops. That is because
it might be the case that $\intpoly{\transitions}$ has a \mlrf but
$\transitions$ does not.

\bexm 
\label{ex:ints}
Consider the loop
\[
\begin{array}{l}
\while~( x_2-x_1 \le 0, x_1+x_2 \ge 1, x_3 \ge 0 ) ~\wdo~ x_2' = x_2-2x_1+1;  x_3' = x_3 + 10x_2+9 
\end{array}
\]
When interpreted over the integers, this loop has the \mlrf
$\tuple{10x_2,x_3}$. However, when interpreted over the rationals, the
loop does not even terminate --- consider the point $(\frac{1}{2},
\frac{1}{2}, 0)$.  
\eexm

For \lrfs, completeness for the integer case was achieved by reducing
the problem to the rational case, using the integer hull
$\inthull{\transitions}$~\cite{CookKRW10,Ben-AmramG13jv}.
In fact, it is quite easy to see why this reduction works for \lrfs,
as the requirements that a \lrf has to satisfy are a conjunction of
linear inequalities and if they are satisfied by
$\intpoly{\transitions}$, they will be satisfied by convex combinations of such points, i.e.,
$\inthull{\transitions}$.

Since we have reduced the problem of finding a \mlrf to finding a
nested ranking function, and the requirements from a nested ranking
function are conjunctions of linear inequalities that should be
implied by $\transitions$, it is tempting to assume that this argument applies
also for \mlrfs.
However, to justify the use of nested functions, specifically in proving
Lemma~\ref{lem:mlrf-pos}, we relied on Lemma~\ref{lem:posfuncstr},
which we applied to $\transitions$ (it is quite easy to see that the lemma fails
if instead of quantifying over a polyhedron, one quantifies only on its integer points).
This means that we did not prove that the existence of a
\mlrf for $\intpoly{\transitions}$ implies the existence of a nested
ranking function over $\intpoly{\transitions}$.
A similar observation also holds for the results of
Section~\ref{sec:bms-mlrf}, where we proved that any (weak) \bmsllrf
can be converted to a \mlrf. Those results are valid only for rational
loops, since in the corresponding proofs we used
Lemma~\ref{lem:posfuncstr}.

In this section we show that reduction of the integer case to the rational one, via the integer hull,
does work also
for \mlrfs, and for converting \bmsllrfs to \mlrfs,  thus
extending the result of sections~\ref{sec:ratcase}
and~\ref{sec:bms-mlrf} to integer loops.
We do so by showing that if $\intpoly{\transitions}$ has a \bmsllrf,
then $\inthull{\transitions}$ has a weak \bmsllrf.

\bthm
\label{th:intbms2wbms}
Let $\tuple{f_1,f_2,\dots,f_d}$ be 
a weak \bmsllrf for
$\intpoly{\transitions}$. Then there are constants $c_1,\dots,c_d$
such that $\tuple{f_1+c_1,\dots,f_d+c_d}$ is a weak \bmsllrf for
$\inthull{\transitions}$ (over the rationals).
\ethm

\bprf 
The proof is by induction on $d$. The base case, $d=1$, concerns a
\lrf, and as already mentioned, is trivial (and $c_1=0$).  
For $d>1$, define:
\begin{align}
\transitions'  &= \inthull{\transitions} \cap \{ \vec{x}'' \in \rats^{2n} \mid f_1(\vec x) \le -1 \}, \label{eq:intcase:1}\\
\transitions'' &= \inthull{\transitions} \cap \{ \vec{x}'' \in \rats^{2n} \mid \diff{f_1}(\vec x'') = 0 \} \label{eq:intcase:2}
\end{align}
Note that $\transitions'$ includes only
integer points of $\inthull{\transitions}$ that are not ranked
at the first component, due to violating $f_1(\vec{x}) \ge 0$.  By changing the first component into $f_1+1$, we take care
of points where $-1 < f_1(\vec{x}) < 0$.  Thus we will have that for every integer point  $\vec{x}''\in \transitions$, if it is not
in $\transitions'$, then the first
component is non-negative, and otherwise $\vec{x}''$ is ranked by $\tuple{f_2,\dots,f_d}$.
Similarly $\transitions''$ includes all the integer points of
$\inthull{\transitions}$ that are not ranked by the first
component due to violating $\diff{f_1}(\vec x'') > 0$.  
Note also that $\transitions''$ is integral, since it is a face of
$\inthull{\transitions}$. On the other hand, $\transitions'$ is not necessarily integral,
so we have to distinguish $\inthull{\transitions}'$ from $\transitions'$.
By the induction hypothesis there are 
\begin{itemize}
\renewcommand{\labelitemi}{$\bullet$}
\item $c'_2,\dots,c'_d$ such that
  $\tuple{f_2+c'_2,\dots,f_d+c'_d}$ is a weak \bmsllrf for
  $\inthull{\transitions}'$; and
\item $c''_2,\dots,c''_d$ such that
  $\tuple{f_2+c''_2,\dots,f_d+c''_d}$ is a weak \bmsllrf for
  $\inthull{\transitions}''$.
\end{itemize}
Next we prove that $f_1$ has a lower bound on
$\inthull{\transitions}\setminus\inthull{\transitions}'$, i.e., there
is a constant $c_1 \ge 1$ such that $f_1+c_1$ is non-negative on this set.
Before proceeding to the proof, note that
this implies that
\[
\tuple{f_1+c_1, f_2+\max(c'_2,c''_2),\dots,f_d+\max(c'_d,c''_d)}
\]
is a weak \bmsllrf for $\inthull{\transitions}$. To see this, take any rational
$\vec{x}''\in\inthull{\transitions}$, then either $\vec{x}''$ is ranked by the first component,
or $\vec{x}''\in\transitions''$
or $\vec{x}''\in\inthull{\transitions}'$; in the last two cases, it is ranked by
a component $f_i+\max(c'_i,c''_i)$ for $i>1$.

It remains to prove that $f_1$ has a lower bound
on $\inthull{\transitions}\setminus\inthull{\transitions}'$.
We assume that $\inthull{\transitions}'$ is non-empty, since otherwise, by the definition of $\transitions'$, it is easy to see
that $f_1 \ge -1$ over all of $\inthull{\transitions}$.
 Thus, we consider $\inthull{\transitions}'$ in
an irredundant constraint representation:
\begin{equation}
\inthull{\transitions}' = \{ \vec{x}'' \in \rats^{2n} \mid \vect{a}_i \cdot \vec{x}'' \le b_i, \ i = 1,\dots,m \},
\end{equation}
and define
\begin{align}
\poly{P}_i &= \inthull{\transitions} \cap \{ \vec{x}'' \in \rats^{2n} \mid \vect{a}_i \cdot \vec{x}'' > b_i \} \\
\poly{P}'_i &= \inthull{\transitions} \cap \{ \vec{x}'' \in \rats^{2n} \mid \vect{a}_i \cdot \vec{x}'' \ge b_i \}
\end{align}
for $i=1,\dots,m$. Then, clearly,
$\inthull{\transitions}\setminus\inthull{\transitions}' \subseteq
\bigcup_{i=1}^m \poly{P}_i$.  It suffices to prove that $f_1$ has a
lower bound over $\poly{P}_i$, for every $i$. Fix $i$ for the rest of
the argument, such that $\poly{P}_i$ is not empty.
It is important to note that, by construction, all integer points of
$\poly{P}_i$ are also in
$\inthull{\transitions}\setminus\inthull{\transitions}'$.

Let $H$ be the half-space $\{\vec{x}'' \!\mid \vect{a}_i\cdot\vec{x} \le
b_i\}$.  We first claim that $\poly{P}_i = \poly{P}'_i \setminus H$
contains an integer point.  Equivalently, there is an integer point of
$\inthull{\transitions}$ not contained in $H$. 
There has to be such a point, for otherwise, $\inthull{\transitions}$,
being integral, would be contained in $H$, and $\poly{P}_i$ would be
empty. Let $\vec{x}''_0$ be such a point.

Next, assume (by way of contradiction) that $f_1$ is \emph{not} lower
bounded on $\poly{P}_i$.
Express $f_1$ as $f_1(\vec{x}) = \vect{\lambda}\cdot \vec{x} +
\lambda_0$, then $\vect{\lambda}\cdot \vec{x}$ is not lower bounded
on $\poly{P}_i$ and thus not on $\poly{P}_i'$.
This means that $\poly{P}_i'$ is not a polytope, and thus can be
expressed as $\poly{O} + \poly{C}$, where $\poly{O}$ is a polytope and
$\poly{C}$ is a cone.
It is easy to see that there must be a rational $\vec{y}''\in
\poly{C}$ such that $\vect{\lambda}\cdot \vec{y} < 0$, since
otherwise $f_1$ would be bounded on $\poly{P}'_i$.

For $k\in \ints_+$, consider the point $\vec{x}''_0 +k\vec{y}''$.
Clearly it is in $P_i'$. 
Since $\vec{y}'' \in \poly{C}$, we have $\vect{a}_i\cdot  \vec{y}'' \ge 0$;
Since $\vec{x}''_0 \in \poly{P}_i$, we have $\vect{a}_i\cdot \vec{x}''_0  > b_i$; adding up,
we get $\vect{a}_i\cdot (\vec{x}''_0 + k\vec{y}'') > b_i$ for all $k$.
We conclude that the set
$S = \{\vec{x}''_0 + k\vec{y}'' \mid k\in \ints_+ \}$ is
contained in $\poly{P}_i$. Clearly, it includes an infinite number of integer
points.
Moreover $f_1$ obtains arbitrarily negative values on $S$ (the larger
$k$, the smaller the value), in particular on its integer points.
Recall that these points are included
$\inthull{\transitions}\setminus\inthull{\transitions}'$, thus $f_1$ is not
lower bounded on the \emph{integer points of}
$\inthull{\transitions}\setminus\inthull{\transitions}'$, a contradiction to the
way $\inthull{\transitions}'$ was defined.
\eprf

\bcor
If $\intpoly{\transitions}$ has a weak \bmsllrf of depth $d$, then
$\inthull{\transitions}$ has a \mlrf, of depth at most $d$.
\ecor

\bprf
By  Theorem~\ref{th:intbms2wbms} we know that
$\inthull{\transitions}$ has a weak \bmsllrf (of the same depth), which in turn can be
converted to a \mlrf by Theorem~\ref{thm:wbms2mlrfQ}.
\eprf

Since \mlrfs are
also weak \bmsllrf, we get

\bcor
If $\intpoly{\transitions}$ has a \mlrf of depth $d$, then $\inthull{\transitions}$
has a \mlrf of depth at most $d$.
\ecor

The above corollary provides a complete procedure for synthesizing
\mlrfs over the integers, simply by seeking a nested ranking function
for $\inthull{\transitions}$.

\bexm
\label{ex:ints:ih}
For the loop of Example~\ref{ex:ints}, computation of the integer hull
results in the addition of the constraint $x_1\ge 1$. Now seeking a
\mlrf as in Section~\ref{sec:ratcase} we find, for example,
$\tuple{10x_2+10,x_3}$.  Note that $\tuple{10x_2,x_3}$, which a \mlrf for
$\intpoly{\transitions}$, is not a \mlrf for $\inthull{Q}$ according
to Definition~\ref{def:mlrf}, e.g., for any $0<\eps<1$ the transition
$(1+\eps,-\eps,0,1,-3\eps-1,-10\eps+9) \in \inthull{Q}$ is not ranked,
since $10x_2<0$ and $x_3-x_3' = 10\eps-9 < 1$.
\eexm

The procedure described above has exponential-time complexity,
because computing the integer hull requires exponential time. However,
it is polynomial for the cases in which the integer hull can be
computed in polynomial time~\cite[Sect.~4]{Ben-AmramG13jv}.
The next theorem shows that the exponential time complexity is
unavoidable for the general case (unless $\mathtt{P}=\NP$).

\bthm 
\label{thm:coNP-complete}
$\dmlrfz$ is \coNP-complete.
\ethm

The proof repeats the arguments in the \coNP-completes proof for linear
ranking functions~\cite[Sect.~3]{Ben-AmramG13jv}. We omit the details.

%% file: depth-bound.tex

\section{The Depth of a \mlrf}
\label{sec:depth}

A wishful thought: If we could pre-compute an upper bound on the depth
of optimal \mlrfs, and use it to bound the recursion, we would obtain
a complete decision procedure for \mlrfs in general, since we can seek
a \mlrf, as in Section~\ref{sec:ratcase}, of this specific depth.
This thought is motivated by results for \emph{lexicographic ranking
  functions}, for example, \cite{Ben-AmramG13jv} shows that the number
of components in such functions is bounded by the number of variables
in the loop.
For \mlrfs, we were not able to find a similar upper bound, and we can show
that the problem is more complicated than in the lexicographic case as
a bound, if one exists, must depend not only on the number of
variables or constraints, but also on the values of the coefficients
in the loop constraints.

\bthm
\label{thm:depthbound}
For integer $B>0$, the following loop $\transitions_B$
\[
\verb/while / (x\ge 1,\; y\ge 1,\; x\ge y,\; 2^{B} y\ge  x) \verb/ do / x'=2x,\; y'=3y
\]
needs at least $B+1$ components in any \mlrf.
\ethm

\bprf
Define $R_i=\{ (2^ic,c,2^{i+1}c,3c) \mid c \ge 1 \}$ and note that for
$i=0\ldots B$, we have $R_i \subset \transitions_B$. Moreover, $R_i
\neq R_j$ for different $i$ and $j$.
Next we prove that
in any \mlrf $\tuple{f_1,\ldots,f_d}$ for $\transitions_B$, and $R_i$
with $i=0\ldots B$, there must be a component $f_k$ such that
$\vec{x}''\in R_i \rightarrow  f_k(\vec{x}) - f(0,0) \ge 0 \land\diff{f_k}(\vec{x}'')>0.$
To prove this, fix $i$.  We argue by the pigeonhole principle that, for some $k$,
$f_k(2^ic,c) = c f_k(2^i,1) + (1-c)f_k(0,0)\ge 0$ and
$f_k(2^ic,c)-f_k(2^{i+1}c,3c)=c(f_k(2^i,1)-f_k(2^{i+1},3))>0$ for
infinite number of values of $c$, and thus $f_k(2^i,1) - f_k(0,0) \ge 0$, and
$f_k(2^i,1)-f_k(2^{i+1},3)>0$, leading to the above statement.
We say that $R_i$ is ``ranked'' by $f_k$.

If $d < B+1$, then, by the
pigeonhole principle, there are different $R_i$ and $R_j$ that are
``ranked'' by the same $f_k$. We show that this leads to a contradiction.
Consider $R_i$ and $R_j$, with $j>i$, and assume that they are
``ranked'' by $f_k(x,y)=a_1x+a_2y+a_0$.
Applying the conclusion of the last paragraph to $R_i$ and $R_j$, we have:
\begin{alignat}{2}
f_k(2^i,1)-f_k(2^{i+1},3) &=  \: -a_12^i-a_22  &\ & > 0 \label{eq:b:dec:1}\\
f_k(2^j,1)-f_k(2^{j+1},3) &=  \: -a_12^j-a_22 && > 0 \label{eq:b:dec:2}\\
f_k(2^i,1) - f_k(0,0) &=   a_12^i+a_2  && \ge 0 \label{eq:b:pos:1}\\
f_k(2^j,1) - f_k(0,0) &=   a_12^j+a_2  && \ge 0 \label{eq:b:pos:2}
\end{alignat}
Adding $\frac{1}{2}\cdot$\eqref{eq:b:dec:2}  to
\eqref{eq:b:pos:2} we get
\highlight{gray!10}{$a_1 2^{j-1} > 0$}.
Thus, $a_1$ must be positive.
From the sum of $\frac{1}{2}\cdot$\eqref{eq:b:dec:2} and \eqref{eq:b:pos:1}, we get
$a_1(2^i-2^{j-1}) > 0$, which implies $j>i+1$, and $a_1<0$, a contradiction. We conclude that $d$
must be at least $B+1$.
\eprf

The bound $B+1$ in the above theorem is tight. This is
confirmed by the \mlrf $\tuple{x-2^By , x-2^{B-1}y , x-2^{B-2}y ,
  \ldots, x-y}$. 

%% file: iter-bound.tex

\section{Iteration Bounds from \mlrfs}
\label{sec:iter-bound}

Automatic complexity analysis techniques are often based on bounding
the number of iterations of loops, using ranking functions.  Thus, it
is natural to ask if a \mlrf implies a bound on the number of
iterations of a given \slc loop.
For \lrfs, the implied bound is trivially linear, and
in the case of \slc loops, it is
known to be linear also for a class of lexicographic ranking functions~\cite{Ben-AmramG13jv}. In this section we
show that \mlrfs, too, imply a linear iteration bound, despite the fact that the variables involved may grow non-linearly during
the loop.

\bexm
\label{ex:iterbound:1}
Consider the following loop
\[
\verb/while / (x\ge 0) \verb/ do / x'=x+y,\; y'=y-1
\]
with the corresponding \mlrf $\tuple{y+1,x}$. 
Let us consider an execution starting from positive values $x_0$ and
$y_0$. It is easy to see that when $y+1$ reaches $0$, i.e., when
moving from that first phase to the second phase, the value of $x$
would be $x_0+y_0+(y_0-1)+\ldots+1+0-1=x_0+\frac{y_0(y_0+1)}{2}-1$,
which is polynomial in the input. It may seem that the
next phase would be super-linear, since the second phase 
is ranked by $x$, however, note that $x$ decreases first by $1$,
then by $2$, then by $3$, and so on. This means that the quantity
$\frac{y_0(y_0+1)}{2}$ is eliminated in $y_0$ iterations, and then
in at most $x_0-1$ iteration the remaining value $x_0-1$ is
eliminated as well. Thus, the total runtime is linear.
\eexm

In what follows we generalize the observation of the above example. We
consider an \slc loop $\transitions$, and a corresponding irredundant
\mlrf $\tau=\tuple{f_1,\dots,f_d}$.
Let us start with an outline of the proof. We first define a function
$F_k(t)$ that corresponds to the value of $f_k$ after iteration $t$.
We then bound each $F_k$ by some expression $\emph{UB}_k(t)$, and
observe that for $t$ greater than some number $T_k$, that depends
linearly on the input, $\emph{UB}_k(T_k)$ becomes
negative. This means that $T_k$ is an upper bound on the time in which
the $k$-th phase ends; the whole loop must terminate before $\max_k
T_k$ iterations.

Let $\vec{x}_t$ be the state after iteration $t$, and define $F_k(t)
= f_k(\vec x_t)$, i.e., the value of the $k$-th component $f_k$ after
$t$ iterations. For the initial state $\vec{x_0}$, we let $M =
\max(f_1(\vec{x_0}),\dots,f_d(\vec{x_0}),1)$.
Note that $M$ is linear in
$\Vert \vec x_0 \Vert_{\infty}$ (i.e., in the maximum absolute value
of the components of $\vec{x}_0$).
We first state an auxiliary lemma and then a lemma that bounds $F_k$.

\blem
\label{lem:use-posfunclem}
For all $1 < k \le d$, there are $\mu_1,\dots,\mu_{k-2} \ge 0$ and
$\mu_{k-1} > 0$ such that
$\vec{x}''\in\transitions \:\rightarrow  \mu_1  f_1(\vec x) + \dots + \mu_{k-1}  f_{k-1}(\vec x) + (\diff{f_k}(\vec x'')-1) \ge 0$.
\elem

\bprf
From the definition of \mlrf we have
$$\vec{x}''\in\transitions\rightarrow  f_1(\vec x) \ge 0 \vee \cdots \vee f_{k-1}(\vec x) \ge 0 \vee \diff{f_k}(\vec x'')  \ge 1.$$
Moreover the conditions of Lemma~\ref{lem:posfunc} hold since $f_k$ is
not redundant, thus there are non-negative constants
$\mu_1,\ldots,\mu_{k-1}$ such that
$$\vec{x}''\in\transitions \rightarrow \mu_1 f_1(\vec{x})+\cdots+
\mu_{k-1} f_{k-1}(\vec{x})+ (\diff{f_{k}}(\vec{x}'') - 1) \geq
0.$$
Moreover, at least $\mu_{k-1}$ must be non-zero, otherwise it means
that $\diff{f_k}(\vec{x}'')\ge 1$ holds already when
$f_1,\dots,f_{k-2}$ are negative, so $f_{k-1}$ would be redundant.
\eprf

It is not hard to see that the coefficients described in the above lemma can be computed explicitly,
if desired. Similarly, the constants in the next lemma, and consequently the linear iteration bound we claim,
can be computed explicitly, in polynomial time.

\blem
For all $1\le k\le d$, there are constants $c_k, d_k > 0$ such that $
F_k(t) \le c_k M t^{k - 1} - d_k t^{k}$, for all $t \ge 1$.
\elem

\bprf 
The proof is by induction. For the base case, $k=1$, we take
$c_1=d_1=1$ and get $F_1(t) \le M - t$, which is trivially true.
For $k\ge 2$ we assume that the lemma holds for smaller indexes and
show that it holds for $k$.  First note that the change in the value
of $F_k(t)$ in the $i$-th iteration is
$f_k(\vec{x}_{i+1})-f_k(\vec{x}_{i}) = -\diff{f_k}(\vec{x}''_i)$.
By Lemma~\ref{lem:use-posfunclem} and the definition of $F_k$,
we have $\mu_1,\dots,\mu_{k-2} \ge 0$ and $\mu_{k-1}>0$ such that
(over $\transitions$)
\begin{equation}
\label{eq:ib:0}
f_k(\vec{x}_{i+1}) - f_k(\vec{x}_{i})  <  \mu_1  F_1(i) + \dots + \mu_{k-1}  F_{k-1}(i)  \,.
\end{equation}
Now we bound $F_k$ (explanation follows):
\begin{align}
F_k(t)  &= f_k(\vec{x}_0) + \Sigma_{i=0}^{t-1} (f_k(\vec{x}_{i+1}) - f_k(\vec{x}_{i})) \label{eq:ib:1}
\\ & < M + \Sigma_{i=0}^{t-1} \left( \mu_1  F_1(i) + \dots + \mu_{k-1}  F_{k-1}(i) \right) \label{eq:ib:2}
\\ & \le M(1+\mu) + \Sigma_{i=1}^{t-1} \left( \mu_1  F_1(i) + \dots + \mu_{k-1}  F_{k-1}(i)  \right) \label{eq:ib:3}
\\ & \le  M(1+\mu) +   \Sigma_{i=1}^{t-1}   \Sigma_{j=1}^{k-1} \left( \mu_j c_j M i^{j - 1}  - \mu_j d_j i^{j} \right)  \label{eq:ib:4}
\\ & \le  M(1+\mu) +   \Sigma_{i=1}^{t-1}  \left ( ( \Sigma_{j=1}^{k-1} \mu_j c_j M i^{j - 1}  )   - \mu_{k-1} d_{k-1} i^{k-1} \right) \label{eq:ib:5} 
\\ & \le  M(1+\mu) +   \Sigma_{i=1}^{t-1}  \left ( M ( \Sigma_{j=1}^{k-1} \mu_j c_j ) i^{k - 2}  -  \mu_{k-1} d_{k-1} i^{k-1} \right) \label{eq:ib:6}
\\ & =  M(1+\mu) +   M ( \Sigma_{j=1}^{k-1} \mu_j c_j ) (  \Sigma_{i=1}^{t-1}  i^{k - 2}  )  -   \mu_{k-1} d_{k-1}\Sigma_{i=1}^{t-1} i^{k-1} \label{eq:ib:7}
\\ & \le  M(1+\mu) +   M ( \Sigma_{j=1}^{k-1} \mu_j c_j ) (  \frac{t^{k-1}}{k-1}  )  -   \mu_{k-1} d_{k-1} ( \frac{t^{k}}{k} - t^{k-1} ) \label{eq:ib:8}
\\ & =  c_k M  t^{k-1}   -   d_k t^{k} \label{eq:ib:9}
\end{align}
Each step above is obtained from the previous one as follows:
\eqref{eq:ib:2} by replacing $f_k(\vec{x}_0)$ by $M$, since
$f_k(\vec{x}_0)\le M$, and applying~\eqref{eq:ib:0};
\eqref{eq:ib:3} by separating the term for $i=0$ from the sum; this
term is bounded by $\mu M$, where $\mu=\sum_{j=1}^{k-1}\mu_j$, because
$F_k(0)=f_k(\vec{x}_0)\le M$ by definition;
\eqref{eq:ib:4} by applying the induction hypothesis;
\eqref{eq:ib:5} by removing all negative values $ - \mu_j d_j i^{j}$, except the last one
$-\mu_{k-1} d_{k-1} i^{k-1}$;
\eqref{eq:ib:6} by replacing $i^{j-1}$ by an upper bound $i^{k-2}$;
\eqref{eq:ib:7} by opening parentheses;
\eqref{eq:ib:8} replacing $\Sigma_{i=1}^{t-1} i^{k - 2}$ by an
upper bound $\frac{t^{k-1}}{k-1}$, and $\Sigma_{i=1}^{t-1} i^{k-1}$ by
a lower bound $ \frac{t^{k}}{k} - t^{k-1}$;
finally for \eqref{eq:ib:8}, take $c_k=(1+\mu)+(\Sigma_{j=1}^{k-1}
\mu_j c_j)/(k-1)+\mu_{k-1} d_{k-1}$, and $d_k=\mu_{k-1} d_{k-1}/k$ and
note that both are positive.
\eprf

\bthm
An \slc loop that has a \mlrf terminates in a number of iterations
bounded by $O(\Vert \vec x_0 \Vert_{\infty})$.
\ethm

\bprf
Observe that for $t > \max\{1, (c_k/d_k)M\}$, we have $F_k(t) < 0$,
proving that the $k$-th phase terminates by this time (since it
remains negative after that time). Thus, by the time
$\max\{1,(c_1/d_1)M,\ldots,(c_k/d_k)M\}$,
which is linear in $\Vert \vec x_0 \Vert_{\infty}$ since $M$ is, all
phases must have terminated.
\eprf

Note that the constants $c_k$ and $d_k$ above can be computed explicitly, if desired.

\bexm 
\label{ex:iterbound:2}
Consider the loop of Example~\ref{ex:iterbound:1}, the corresponding
\mlrf $\tuple{y+1,x}$, initial positive input $y_0 \ge x_0 \ge 1$, and
let $M=\max(x_0,y_0+1)=y_0+1$.
By definition $F_1(t) \le M-t$. Let us bound $F_2$. First we find a
bound on $-\diff{f_2}(\vec{x}'')$. Starting from
\[
(x,x',y,y') \in \transitions \rightarrow y+1\ge 0 \vee (x-x') \ge 1
\]
we find $\mu_1>0$ such that
\[
(x,x',y,y') \in \transitions \rightarrow \mu_1(y+1)+(x-x')-1 \ge 0
\]
which holds for $\mu_1=1$ (substitution $x'=x+y$ and $\mu_1=1$ we get
$0\ge 0$). Thus $-\diff{f_2}(\vec{x}'') = x'-x \le y$ (which is easy
to see in this case since the update is a deterministic). Now $F_2(t)
\le c_2 M t - d_2 t^{2}$ where $c_2=(1+\mu)+(\mu_1 c_1)+\mu_{1}
d_{1}=4$, and $d_k=\mu_{1} d_{1}/2=\frac{1}{2}$. Thus $F_2(t) < 4 M t
- \frac{1}{2} t^{2}$.
For $t>\max\{1,M,8M\}$ we get both $F_1(t)$ and $F_2(t)$ negative,
which means that $8M=8y_0+8$ is a bound for the runtime for this
input.
\eexm

We remark that the above result also holds for multi-path loops if they have a nested ranking function,
but does not hold for any \mlrf.

%% file: conclusions.tex

\section{Conclusion}
\label{sec:conc}

Linear ranking functions, lexicographic and multiphase combinations of linear ranking functions, have all been proposed in earlier work.
The original purpose of this work has been to improve our understanding of multiphase functions, and answer open problems
regarding the complexity of obtaining such
functions from linear-constraint loops, the difference between the integer case and the rational case, and the possibility of inferring 
an iteration bound from such ranking functions. 
Similarly, we wanted to understand a natural class of lexicographic ranking functions, which removes a restriction of previous definitions regarding
negative values.
Surprisingly, it turned out that our main results are \emph{equivalences} which show
that, for single-path linear-constraint loops, both \mlrfs and \llrfs reduce to a simple kind of \mlrf,
that has been known to allow polynomial-time solution (over the rationals).  
Thus, our result collapsed, in essence, the above classes of ranking functions.

The implication of having a polynomial-time solution, which is hardly
more complex than the standard algorithm to find linear ranking functions, is that whenever one considers using \lrfs in one's work, one should consider
using \mlrfs. By controlling the depth of the \mlrfs one trades expressivity for processing cost.  We believe that it would be sensible to start with depth 1
(i.e., seeking a \lrf) and increase the depth upon failure. Similarly, since a complete solution for the integers is inherently more costly (as we proved it
to be \coNP-complete), it makes sense to begin with the solution that is complete over the rationals, since it is, at any rate, safe for the integer case.
If this does not work, one can also consider special cases in which the inherent hardness can be avoided, as discussed in detail in~\cite[Sect.~4]{Ben-AmramG13jv}.

Theoretically, some tantalizing open problems remain. Is it possible to decide whether a given loop admits a \mlrf, without a depth bound?
This is related to the question, discussed in Section~\ref{sec:depth}, whether it is possible to precompute a depth bound.
What is the complexity of the \mlrf problems over multi-path loops? For such loops, the equivalence \mlrfs, nested r.f.s and \llrfs does not hold.
Finally (generalizing the first question), we think that there is need for further exploration of single-path loops and of 
the plethora of ``termination witnesses" based on linear functions (a notable reference is~\cite{LeikeHeizmann15}).

We have implemented the \emph{nested ranking function} procedure of
Section~\ref{sec:ratcase}, and applied it, among others, on a set of
terminating and non-terminating \slc loops taken from~\cite{GantyG13}.
These examples originate mainly from~\cite{ChenFM15}, and they were
collected as ones that require the transition invariants
techniques~\cite{DBLP:conf/lics/PodelskiR04} for proving termination.
For all $25$ terminating loops in this set we found a \mlrf ($2$ have
also a \bmsllrf as defined in~\cite{Ben-AmramG13jv} and $6$
have \lrf).
The implementation can be tried at
\url{http://loopkiller.com/irankfinder}, where this set of examples is
available as well.

Closely related work is already discussed in Section~\ref{sec:intro},
for more details on algorithmic and complexity aspect of linear
ranking of \slc loops, we refer the reader to~\cite{Ben-AmramG13jv}.